\documentclass[
  reprint,              
  amsmath,              
  amssymb,              
  aps,                  
  prd,                  
  superscriptaddress,   
  floatfix              
]{revtex4-2}

\usepackage{graphicx}                   
\usepackage[caption = false]{subfig}
\usepackage{dcolumn}                    
\usepackage{bm}                         
\usepackage[colorlinks=true,
            citecolor=blue,
            linkcolor=blue,
            urlcolor=blue]{hyperref}    
\usepackage{physics}                    
\usepackage{xcolor}                     
\usepackage{amsmath}
\usepackage{booktabs}

\usepackage{tabularray}
\UseTblrLibrary{booktabs}

\usepackage{tikz}
\usetikzlibrary{positioning, matrix, decorations.markings}
\usetikzlibrary{calc}

\begin{document}

\title{Some Mechanisms for the Quantum Mpemba Effect}

\author{Rodrigo F. Saliba}
\affiliation{Departamento de Física, Universidade Federal de Minas Gerais,}
\author{Vitor M. Pereira}
\affiliation{Departamento de Física, Universidade Federal de Minas Gerais,}
\author{Raphael C. Drumond}
\affiliation{Departamento de Matemática, Universidade Federal de Minas Gerais}

\begin{abstract}
\textit{In recent years, the quantum Mpemba effect (QME) — which occurs when an out-of-equilibrium system reaches equilibrium faster than another that is closer to equilibrium — has attracted significant attention from the scientific community as an intriguing and counterintuitive phenomenon. It generalizes its classical counterpart, extending the concept beyond thermal equilibration. This paper approaches the QME for Markovian open quantum systems from different perspectives. Firstly, we propose a physical mechanism based on decoherence-free subspaces. Secondly, we show that a macroscopic enhancement of the decay rate that scales linearly with system size can be obtained, leading to an extreme version of the phenomenon on Markovian open quantum systems. Thirdly, we study the Mpemba effect through the unravelings of Davies maps, revealing subtleties in the figures of merit which are chosen for obtaining the QME. Lastly, we propose a microscopic model for a better understanding of thremal bath dynamics in this context.}  
\end{abstract}

\date{\today}
\maketitle

\section{Introduction}

Throughout history several cultures, philosophers, and scientists have written about the phenomenon in which an initially hot system cools down faster than an initially colder one. Some of the historical figures who noticed and described this effect include Aristotle~\cite{ross1923works}, Francis Bacon~\cite{bacon2011works}, and René Descartes~\cite{descartes1824discours}. The first systematic studies of the phenomenon were conducted by Mpemba, Osborne, and Kell~\cite{mpemba1969cool,kell1969freezing}, and the effect has since been referred to as the \textit{Mpemba effect}.

Numerous studies have explored the Mpemba effect in different types of systems since Mpemba and Osborne’s initial work~\cite{chaddah2010overtaking,hu2018conformation,greaney2011mpemba}, although the existence of the effect as originally described by Mpemba remains a topic of debate in the literature~\cite{burridge2016questioning,bechhoefer2021fresh}.

Nevertheless, descriptions of quantum analogs of this classical effect are now well established~\cite{chatterjee2023quantum,chatterjee2024multiple,bao2026initial,summer2026resource}. These Mpemba-like effects occur when an initial out-of-equilibrium state reaches equilibrium faster than a state closer to equilibrium. This phenomenon, known as the \textit{quantum Mpemba effect} (QME), has been observed in both closed and open quantum systems~\cite{ares2025quantum}.

In the context of Markovian open quantum systems, recent work has shown that for systems governed by specific types of equations of motion, it is always possible to find a unitary transformation of an initial state that leads to a state with higher energy, which in turn exhibits an exponentially faster relaxation toward equilibrium given an arbitrary initial state with coherences in the energy eigenbasis~\cite{moroder2024thermodynamics,carollo2021exponentially}.

Although mathematically rigorous frameworks for obtaining such states have been developed, the physical mechanisms underlying QME in Markovian open quantum systems are still poorly understood~\cite{longhi2025quantum,longhi2025quantum2}. Here, we explore one possible mechanism and its consequences. If the dynamics of the system has an steady state and is such that the dissipator may be broken into two parts, one of which has a decoherence-free subspace (DFS)~\cite{lidar1998decoherence, lidar2003decoherence}, than it may be possible to find a pair of states, one in the DFS and another outside it which exhibit the QME. This possibility arises, essentially, from the fact that part of the dissipator does not affect the state in the DFS.
Moreover, we study the previously explored method for obtaining states exhibiting the quantum Mpemba effect~\cite{moroder2024thermodynamics,summer2026resource} through the lens of quantum trajectories, aiming for a deeper understanding of the phenomenon in this context. Finally, we approach the quantum Mpemba effect through the lens of a microscopic model where we keep track of system and bath states.

The paper is organized as follows: Section~\ref{section2} reviews the mathematical description of the quantum Mpemba effect. Section~\ref{section3} presents one possible physical mechanism for the QME. Section~\ref{section4} discusses the quantum Mpemba effect through the lens of quantum trajectories. Section~\ref{section5} studies system--bath interactions with the help of a microscopic model based on Gaussian systems.

\section{Mpemba effect on Markovian Open Quantum Systems}
\label{section2}

Markovian open quantum system dynamics obey an equation of motion that can, in general, be written in the Lindblad form~\cite{breuer2002theory,lindblad1976generators,gorini1976completely}:
\begin{equation}
    \frac{d\rho}{dt} = \mathcal{L}[\rho]= -i\left[H,\rho\right]+\mathcal{D}(\rho),
    \label{LindbladEq}
\end{equation}
where
\begin{equation}
\mathcal{D}(\rho)=\sum_k \lambda_k\left( L_k\rho L_{k}^{\dagger} - \frac{1}{2} \left\{L_k^{\dagger}L_k,\rho\right\}\right).
\end{equation}
Here, $\mathcal{L}$ is the Liouvillian, $H$ the Hamiltonian, $\mathcal{D}(.)$ the \emph{dissipator}, $\lambda_k$ are the dissipation rates, $[\cdot,\cdot]$ and $\{\cdot,\cdot\}$ denote the commutator and anticommutator, respectively, and $L_k$ are \textit{jump operators}. 

An important class of such equations are the so called \textit{Davies maps}~\cite{davies1974markovian}, with the important property that their unique steady state is the thermal (Gibbs) state,
\begin{equation}
\rho_{ss} = \frac{e^{-\beta H}}{\mathrm{Tr}[e^{-\beta H}]}.
\label{thermalstate}
\end{equation}
This allows for the observation of the quantum Mpemba effect through thermalization, in close analogy to its classical counterpart.

There are many quantities that can be used to measure the distance between two states~\cite{ares2025quantum}. Here, we focus on  two of those. The first is the trace distance, which captures the optimal distinguishability between two states with a single measurement:
\begin{equation}
    d_{\text{Tr}}(\rho,\rho_{ss}) = \frac{1}{2}\mathrm{Tr}\left[\sqrt{A^{\dagger} A}\right]
\end{equation}
where $A = \rho - \rho_{ss}$. This metric is monotonic in time under Markovian dynamics, which is a desired property in this setup. Here the QME will occur if there are states $\rho(0)$, $\rho'(0) = U\rho(0)U^{\dagger}$, such that $d_{\mathrm{Tr}}(\rho'(0),\rho_{ss}) > d_{\mathrm{Tr}}(\rho(0),\rho_{ss})$ and a time $\tau$ for which $d_{\mathrm{Tr}}(\rho(t),\rho_{ss}) > d_{\mathrm{Tr}}(\rho'(t),\rho_{ss}) > 0 \text{ } \forall t> \tau$. 

The second quantity used to observe the QME in this work is the non-equilibrium free energy, following the approach  of Moroder et al.~\cite{moroder2024thermodynamics}:
\begin{equation}
    F_{neq} = \langle H \rangle - TS_{vn},
    \label{Fneq}
\end{equation} 
where $H$ is the system Hamiltonian, $T = \frac{1}{\beta}$ is the temperature of the equilibrium (thermal) state (see Eq.~\ref{thermalstate}) and $S_{vn} = -\mathrm{Tr}\left[\rho \ln\rho \right]$. 
Although $F_{neq}$ is not itself a metric, it has desirable properties which allow the observation of QME on Davies maps in a viewpoint that is closer to that of the classical Mpemba Effect.  Firstly, it can be written as 
\begin{equation}
    F_{neq} = T D(\rho(t) \vert\vert \rho_{ss}) + F_{eq}(\rho_{ss}),
\end{equation}
where $D(\chi \vert\vert \xi) = \mathrm{Tr}[\chi (\mathrm{ln}\chi - \mathrm{ln} \xi)]$ is the quantum relative entropy and $F_{eq}(\xi) = -T \mathrm{ln}(Z)$, with $Z = \mathrm{Tr}(e^{-\beta H})$. 
Secondly, it also decays monotonically in time and $F_{neq}(\rho(t)) \geq F_{eq}(\rho_{ss}) \text{ } \forall t$ (Klein inequality). Thus, QME occurs if there is a time $\tau$ such that $F_{neq}(\rho'(t)) < F_{neq}(\rho(t))$ for $t > \tau$, where the initial state $\rho'$ has a higher non-equilibrium free energy than $\rho$.
A similar behavior for both quantities, then, is expected, with the exception of the regime where the bath is at zero temperature (see Section~\ref{section4}).




A mathematical pathway for obtaining the QME for Markovian open quantum systems consists of expanding the state of the system in terms of the eigenoperators of the Liouvillian:
\begin{equation}
    \rho(t) = e^{\mathcal{L}t}\rho(0) = \rho_{ss} + \sum_{k = 2}^{D^{2}}\mathrm{Tr}[\mathbf{l}_{k}\rho(0)]\mathbf{r}_k e^{\lambda_kt},
\end{equation}
%
%
where $D$ is the dimension of the Hilbert space, $\rho_{ss}$ is the steady state (this is the right eigenoperator of $\mathcal{L}$ associated with the zero eigenvalue and, as mentioned, thermal under Davies maps), $\mathbf{l}_k$ and $\mathbf{r}_k$ are the left and right eigenoperators of $\mathcal{L}$ associated with the eigenvalue $\lambda_k$, respectively \footnote{$\mathcal{L}[\mathbf{r}_k] = \lambda_{k}\mathbf{r}_k$ and $\mathcal{L}^{\dagger}[ \mathbf{l}_k ]= \lambda_k\mathbf{l}_k$}.
Given an initial state $\rho(0)$, if the eigenvalues are ordered in ascending order according to absolute value of their real parts, i.e.,
\[
0 = \lambda_1 < \lvert \mathrm{Re}(\lambda_2) \rvert \leq \lvert \mathrm{Re}(\lambda_3) \rvert \leq \dots,
\]
assuming a spectral gap given by $\lvert \mathrm{Re}(\lambda_2) \rvert>0$ , it can be shown that an exponential speed-up of system thermalization can be achieved if there exists a unitary operator \(U\) acting on the initial state $\rho(0)$ \(\mathbf{l}_2\) such that 
\(\mathrm{Tr}\!\left[\mathbf{l}_2\, U \rho(0) U^{\dagger}\right] = 0\)~\cite{carollo2021exponentially},
where it is further assumed that \(\lambda_2\) is real and that the initial state is pure (see Fig. \ref{Mpembafigure1}).

\begin{figure}[htbp]

    \begin{tikzpicture}[font = \scriptsize]
        \begin{scope}[
            thick, blue!60!black, x radius = 3.50 cm, y radius = 3.00 cm,
            decoration={
                markings,
                mark = at position 2 cm with {
                    \node [below left = 4pt, black, font = \large] {$\mathcal{D}$};
                },
            }]

            \draw [postaction={decorate}] (0, 0) circle [];
        \end{scope}

        \coordinate (A) at (270:2.0);
        \fill (A) circle [radius = 1 pt];
        \node [below = 2 pt, font = \normalsize] at (A) {$\rho_{ss}$};
        
        \coordinate (B) at (0, 0);
        \fill (B) circle [radius = 1 pt];
        \node [above right = 4 pt and 20 pt, anchor = south] at (B) {$
            \rho(t) \propto \tilde{\rho}^{\lambda_{2}t} 
            + \mathcal{O}(e^{\lambda_{3}t})
        $};

        \coordinate (C) at (160:2.5);
        \fill (C) circle [radius = 1 pt];
        \node [above left = 4 pt and 10 pt, anchor = south west] at (C) {$
            \rho'(t) \propto \tilde{\rho}e^{\lambda_{3}t} 
            + \mathcal{O}(e^{\lambda_{4}t})
        $};

        \draw [->, shorten >= 3pt] (B) .. controls +(180:2) and +(170:2) .. (270:0.75)
                                       .. controls +(350:4) and +(0:1) .. (A);

        \draw [->, shorten >= 3pt] (C) .. controls +(260:2) and +(170:1) .. (A);

        \coordinate (L1) at ($(C) + (-1.35, 0)$);
        \coordinate (L2) at (B -| L1);
        \coordinate (L3) at (A -| L1);

        \coordinate (M1) at ($(L2) + (-0.05, 0)$);
        \coordinate (M2) at (A -| M1);

        \begin{scope}[dashed, opacity = 0.3]
            \draw [shorten <= -2 pt] (L1) -- (C);
            \draw [shorten <= -2 pt] (L2) -- (B);
            \draw [shorten <= -2 pt] (L3) -- (A);
        \end{scope}

        \draw [] (M1) -- (M2);
        \draw [] (L1) -- (L3) node [pos = 0.20, left = 1 pt] {$d(\rho', \rho_{ss})$}
                              node [pos = 0.65, left = 1 pt] {$d(\rho, \rho_{ss})$};

    \end{tikzpicture}
    \caption{Illustrative picture of the quantum Mpemba effect. Here the state $\rho'$ and $\rho$ are represented in the state space $\mathcal{D}$.  The initial distance from the equilibrium state, $\rho_{ss}$, is represented by $d(\rho^{\prime},\rho_{ss}) > d(\rho,\rho_{ss})$. If for the slowest decaying mode $\lambda_2 \in \mathbb{R}$ we have \(\mathrm{Tr}\!\left[\mathbf{l}_2\, U \rho(0) U^{\dagger}\right] = 0\), $\rho'$ reaches $\rho_{ss}$ faster than $\rho$.}
    \label{Mpembafigure1}
\end{figure}

If \(\lambda_2\) is complex, the slowest contributions for the evolution of the density matrix are given by
\begin{equation}
    \begin{aligned}
        \rho(t) = & \rho_{ss} +\\ e^{\mathrm{Re}(\lambda_2)t}
        &\Big(
            \mathrm{Tr}\!\left[\mathbf{l}_2 \rho(0)\right] \mathbf{r}_2 e^{i\,\mathrm{Im}(\lambda_2)t}
            +\\
            & \mathrm{Tr}\!\left[\mathbf{l}_2^{\dagger} \rho(0)\right] \mathbf{r}_2^{\dagger} e^{-i\,\mathrm{Im}(\lambda_2)t}
        \Big) + \mathcal{O}(e^{Re(\lambda_3)}).
    \end{aligned}
\end{equation}

Thus, the unitary operator must satisfy both 
\(\mathrm{Tr}\!\left[\mathbf{l}_2\, U \rho(0) U^{\dagger}\right] = 0\) and 
\(\mathrm{Tr}\!\left[\mathbf{l}_2^{\dagger} U \rho(0) U^{\dagger}\right] = 0\). For Davies maps, a procedure can then be developed to obtain the corresponding unitary for an initial state \(\rho(0)\) with coherences in the energy eigenbasis. This transformation consists in bringing the state to a diagonal form on the same basis~\cite{moroder2024thermodynamics}.

\begin{figure}[htbp]
    \subfloat[]{
        \includegraphics[width=0.75\linewidth]{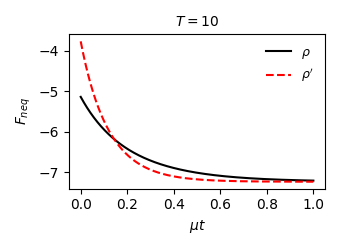}
        \hskip 3em
        \label{fig:examplempembaF}
    }
    \\ 
    \subfloat[]{
        \includegraphics[width=0.75\linewidth]{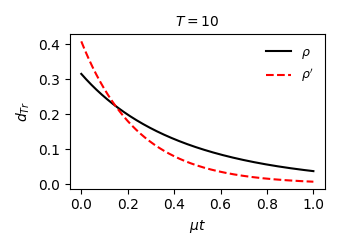}
        \hskip 3em
        \label{fig:examplempembatrd}
    }
    \caption{Evolution of the non-equilibrium free energy \protect\subref{fig:examplempembaF} and trace distance \protect\subref{fig:examplempembatrd} for states $\rho$ and $\rho^{\prime} = U\rho U^{\dagger}$. The reference state is the equilibrium state given by Eq.~\eqref{thermalstate}. The state $\rho^{\prime}$ is seen to have a lower $F_{neq}$ and $d_{\text{Tr}}$ than $\rho$ after a certain time elapse which characterizes the quantum Mpemba effect in this system. Here we used $\omega = 5$, $\mu_+ = \mu_- = \mu = 1$ and $T = 10$ and the initial state was determined by the Bloch vector $\boldsymbol{R} = (0.52807291, 0.21585042, 0.02214326)$.}
 \label{examplempemba2}
\end{figure}

As an example, consider a single qubit coupled to a bosonic bath at temperature \(T\), with Hamiltonian \(H = \frac{1}{2}\omega\sigma_{z}\).
The state of the system can be expressed in terms of the Pauli matrices as
\begin{equation}
\rho = \frac{1}{2}\left[ \mathbb{I} + r_1 \sigma_{x} + r_2 \sigma_{y} + r_3 \sigma_{z} \right],
\end{equation}
where \(\boldsymbol{R} = (r_1, r_2, r_3)\) is the Bloch vector.
The Davies map for this setup is given by
\begin{equation}
    \frac{d\rho}{dt} = -i[H,\rho] + \mu_-(\bar{N} + 1)\mathcal{D}_{\sigma_-}(\rho) + \mu_+(\bar{N})\mathcal{D}_{\sigma_+}(\rho),
\end{equation}
where $\bar{N} = (\exp(\beta \omega) - 1)^{-1}$ is the Bose-Einstein distribution and $\mathcal{D}_{L_k}(\rho) = L_k\rho L_k^{\dagger} - \frac{1}{2}\{L_k^{\dagger}L_k,\rho\}$.
The results for the evolution of \(F_{\mathrm{neq}}\) and $d_{\text{Tr}}(\xi,\rho_{ss}$) with $\xi = \rho$ and $\xi = \rho^{\prime}$ obtained by the method described above are shown in Fig.~\ref{examplempemba2}, where the Bloch vector for $\rho$ was generated randomly.

\section{Collective Decay and Decoherence-Free Subspaces}
\label{section3}

Consider again a general Lindblad equation. All decoherence arising from the loss of information from the system to the reservoir is captured by the dissipator term $\mathcal{D}$. In general, this drives the system from quantum behavior toward classicality. In contexts such as quantum computation, where the system is intended to evolve via unitary gates, such an evolution is undesirable~\cite{nielsen2010quantum}. Extensive research has explored approaches to avoid or mitigate decoherence in the Markovian limit~\cite{landauer1995quantum, unruh1995maintaining, duan1997preserving, zanardi1998dissipation, lidar1998decoherence}, including the use of decoherence-free subspaces. These subspaces consist of states for which the dissipator vanishes, ensuring that the system remains within the subspace and evolves unitarily throughout its dynamics.


The mathematical explanation for the QME given on the last section guarantees an exponential speedup toward equilibrium. Nevertheless, in many-body setups the unitary transformation can be highly nonlocal and, although $\rho$ may in general be easily obtained, the rotated $\rho^{\prime}$ may be a complex entangled state that requires an unrealistic preparation protocol. Here, we are interested in exploring physical mechanisms for the QME where the pair of states which exhibit the phenomenon can be chosen consistently. As we show in this section, this can be achieved by exploring system symmetries rather than using specific transformations based on eingevalues of the Liouvillian. In this pespective, a DFS can be exploited in order to obtain such a mechanism for the QME.

Indeed, assume a system governed by a Lindblad equation with two dissipators, $\mathcal{D}_{dfs}$ and $\mathcal{D}_{per}$, having different rates $\Gamma$ and $\mu$ respectively and where $\Gamma \gg \mu$. If $\mathcal{D}_{dfs}$ has a DFS, but the combined dissipator $\mathcal{D}_{dfs} + \mathcal{D}_{per}$ has a unique steady state, then the states inside the DFS decay slowly (at rate $\mu$), while the states outside it decay rapidly (at rate $\Gamma$). The QME will take place provided that there exists a state outside the DFS that is farther from equilibrium than a state inside it. In the following, we will discuss a concrete example of such a mechanism.



Consider a system of $L$ non-interacting spins with the Hamiltonian:
\begin{equation}
    H = J_z \sum_{i=1}^{L} \frac{1}{2}\sigma^z_i,
    \label{hamext}
\end{equation}
where $J_z$ is the energy gap between states $\ket{\uparrow_i}$ and $\ket{\downarrow_i}$. The system is assumed to interact with a thermal bath, which induces both collective and local dissipation, in such a way that its dynamics is governed by the master equation.
\begin{equation}
    \begin{aligned}
       \frac{d\rho}{dt} &= -i[H,\rho] + \Gamma(1 + \bar{N})\mathcal{D}_{L_c}(\rho) + \Gamma\bar{N}\mathcal{D}_{L_c^{\dagger}}(\rho)\\
       &+  \sum_k \left[\mu(1 + \bar{N})\mathcal{D}_{\sigma^{-}_k}(\rho)+ \mu\bar{N}\mathcal{D}_{ \sigma^{+}_k}(\rho)\right],
    \end{aligned}
    \label{eqm10}
\end{equation} 
where $L_c = \sum_i \sigma^-_i = S^-_c$ is the Lindblad operator of collective decay, and $\Gamma$ and $\mu$ represent the dissipation strengths associated with collective and local decays, respectively. This model is closely related to superradiance~\cite{bloch2008many,Dicke1954,GrossHaroche1982,damanet2019atom,lidar2003decoherence}, where collective interference of emission pathways leads to enhanced cooperative decay.



\begin{figure}[htbp]
    \subfloat[]{
        \includegraphics[width=0.75\linewidth]{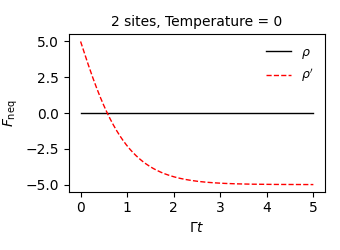}
        \hskip 0.12\linewidth
        \label{fig:3a}
    }
    \\
    \subfloat[]{
        \includegraphics[width=0.75\linewidth]{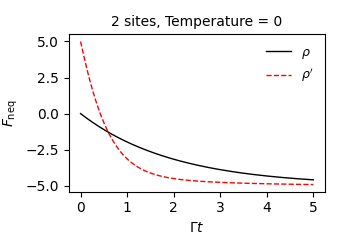}
        \hskip 0.12\linewidth
        \label{fig:subfig21}
    }
    \caption{\protect\subref{fig:3a} Non-equilibrium free energy for triplet (red line) and singlet (black line) states for evolution under collective decay. The result shows that the singlet state is trapped in its initial state if no local dissipator is used ($\mu = 0$). Here, $\Gamma = 1$ and $J_z = 5$. \protect\subref{fig:subfig21} Same parameters, but with $\mu = 0.5$.}
    \label{singletexample}
\end{figure}

Let us first consider the simplest possible case of just two spins, whose only source of dissipation is collective decay at zero temperature, i.e., assume that $L=2$, $\bar{N} = 0$ and $\mu = 0$ in the model. We may take $\rho$ as the pure state defined by $\ket{\psi} = \ket{\uparrow \uparrow}$, which has the maximum energy, while $\rho'$ is the pure state defined by the singlet $\ket{\phi} = \frac{1}{\sqrt{2}}(\ket{\uparrow \downarrow} - \ket{\downarrow \uparrow})$, which has zero initial energy and belongs to the DFS of the collective dissipator. The result of the evolution under the master equation~\eqref{eqm10} for both states is shown in Fig.~\ref{fig:3a}. The singlet state's non-equilibrium free energy is trapped at $F_{neq} = 0$. When local dissipation ($\mu \neq 0$, but with $\mu < \Gamma$) is considered, the singlet state then decays to the steady state, but at a slower rate $\mu$, while the triplet state still decays at a faster rate $\Gamma$. This is shown in Fig.~\ref{fig:subfig21}.  

The QME in this model can easily be generalized to any number $L$ of spins. To this end, we used states from the DFS of the collective dissipator that take the following form:
\begin{equation}
   \ket{\psi} = \frac{1}{\sqrt{L}}\sum_{i = 1}^{L}(-1)^{i}\ket{\downarrow_1}\dots\ket{\downarrow_{i - 1}}\ket{\uparrow_i}\ket{\downarrow_{i+1}}\dots\ket{\downarrow_{L}}.
   \label{decoherence1}
\end{equation}

As a starting point, we evaluated the robustness of this mechanism to temperature changes. Here we considered the initial states, $\rho$ and $\rho'$, as the state in Eq.~\eqref{decoherence1} and the state of all spins up, respectively. We have checked that $\rho'$ reaches the steady state $\rho_{ss}$ faster than $\rho$ and the quantum Mpemba effect still occurs at temperatures greater than zero. 

\begin{figure}[htbp]
    %
    \subfloat[]{
        \includegraphics[width=0.80\linewidth]{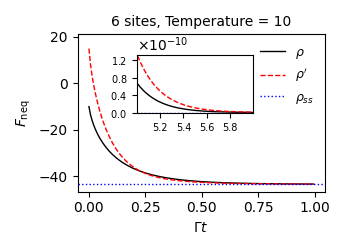}
        \hskip 0.15\linewidth
        \label{fig:thermal10T}
    }
    \\
    \subfloat[]{
        \includegraphics[width=0.80\linewidth]{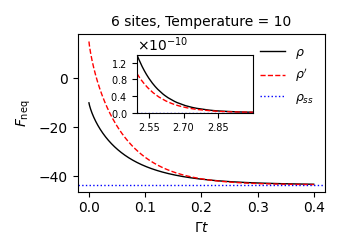}
        \hskip 0.15\linewidth
        \label{fig:thermal10muisgamma}
    }
    \caption{Non-equilibrium free energy for the all-up state $\rho'$ (red line), decoherence-free subspace state $\rho$ (black line) of Eq.\eqref{decoherence1} and steady state $\rho_{ss}$ (blue line) for evolution under collective dissipation at 3 different temperatures, where $T = 0\Gamma, 1\Gamma, \text{ and } 10\Gamma$. For  \protect\subref{fig:thermal10T} the parameters used were $J_z = 5$, $\Gamma = 1$, and $\mu = 0.5$. Although the behavior at low temperatures still characterizes the QME, the definition given in Section \ref{section2} does not hold for $T = 10\Gamma$ which presents 2 crossings in its dynamics. In any case, one can still infer that the second crossing in this case happens when the system is effectively thermalized. \protect\subref{fig:thermal10muisgamma} Shows the behavior for $T=10\Gamma$ and $\mu = \Gamma$. Here we observe only one crossing.}
    \label{extremezerotemperatureexample}
\end{figure}

Interestingly, at $T\gg0$ two different behaviors are observed. These are associated with the order of magnitude of the local dissipation. For $\mu \ll \Gamma$ (Fig.~\ref{fig:thermal10T}) we see that the crossing of the curves occurs twice. On the other hand, if $\mu\approx \Gamma$ (see Fig.~\ref{fig:thermal10muisgamma}) the crossing occurs once, characterizing the QME as defined in Section~\ref{section2}. Nevertheless, the second crossing on the first case occurs at very low differences of $F_{neq}$, where both states are extremely close to the steady state, $\rho_{ss}$. Thus, we may still consider the QME to be valid in this situation, since one can infer that the second crossing happens after the system is effectively thermalized.

\begin{figure}
    \subfloat[]{
         \includegraphics[width=0.75\linewidth]{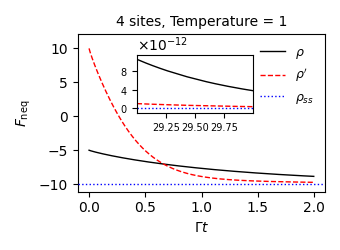}
        \label{fig:subfig11}
    }
    %
    %
    \\
    \subfloat[]{
        \includegraphics[width=0.75\linewidth]{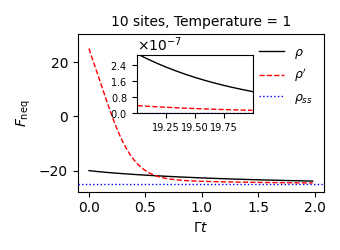}
    }
    \caption{Non-equilibrium free energy for the all-up state $\rho'$ (red line), decoherence-free subspace state $\rho$ (black line) of Eq.~\eqref{decoherence1} and steady state $\rho_{ss}$ (blue line) for evolution under collective dissipation at temperature $T = 1\Gamma$ and different system sizes. Here, the parameters used were $J_z = 5$, $\Gamma = 1$, and $\mu = 0.5$.}
    \label{extremempembat1}
\end{figure}

Next, the behavior of the model was probed with respect to the size of the system. Here we see that the effect still holds, as shown in Fig.~\ref{extremempembat1}. These results provide evidence that one can exploit decoherence-free subspaces as a mechanism to arrive at the quantum Mpemba effect for both few- and many-body systems. 

\subsection{Extreme Quantum Mpemba Effect}

Another consequence of the model shown in Eq.~\eqref{eqm10} can be more easily seen when we compare the all-spins-up state ($\rho'$) with another state from the DFS:

\begin{equation}
   \ket{\psi}_{L} = \bigotimes_{m=1}^{L/2} \ket{s}_{2m-1,2m},
   \label{0initialenergystate}
\end{equation}
where $L$ is even and  $\ket{s}_{i,j} = \frac{1}{\sqrt{2}}(\ket{\uparrow_i}\ket{\downarrow_j} - \ket{\downarrow_i}\ket{\uparrow_j})$. Since this state has zero total magnetization $S_c^{z} = 0$, its initial energy vanishes under the Hamiltonian, independently of the system size.

\begin{figure}
    \subfloat[]{
         \includegraphics[width=0.75\linewidth]{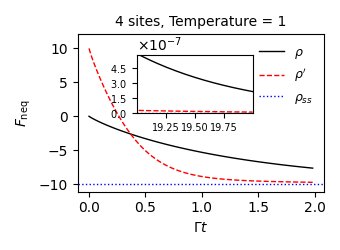}
        \label{fig:subfig113}
    }
    %
    \\
    \subfloat[]{
         \includegraphics[width=0.75\linewidth]{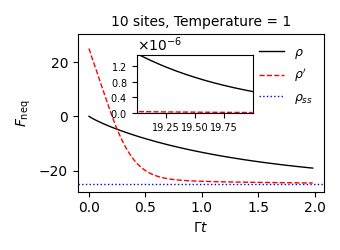}
    }
    \caption{Non-equilibrium free energy evolution for the all-up state, $\rho'$, (red line), decoherence-free subspace state, $\rho$ (black line) of Eq.~\eqref{0initialenergystate} and steady state $\rho_{ss}$ (blue line) for evolution under collective dissipation at temperature $T = 1\Gamma$. Here, the parameters used were $J_z = 5$, $\Gamma = 1$, and $\mu = 0.5$.}
    \label{extremempembatrue}
\end{figure}

From this perspective, the all-up state is seen to decay faster as the system size grows, showing that an extreme version of the quantum Mpemba effect, where the relaxation towards equilibrium of $\rho^{\prime}$ can be made as fast as one desires, can be obtained by increasing the system size. This feature is shown for systems up to 10 sites in Fig.~\ref{extremempembatrue}. Here, the crossing time of the non-equilibrium free energies of both states becomes shorter as the system size grows (see Table~\ref{Table2}).

\begin{table}
    \centering
    \begin{tblr}{
        columns = {6 mm, l, mode=math},
    }
        \toprule
        L         & 4     & 6     & 8     & 10    \\
        \midrule 
        t_{c1}    & 0.37  & 0.32  & 0.28  & 0.25  \\
        \bottomrule
    \end{tblr}
  \caption{Approximate time of crossing ($t_{c1}$) for the results of $F_{neq}$ in Fig.~\ref{extremempembatrue} as the number of sites of the system $L$ grows. Here, $t_{c1}$ indicates the time evolved for temperature set to $T = 1\Gamma$.}
  \label{Table2}
\end{table}
    
In order to show that the result is truly dependent on the system size, we solve the master equation for the simplified model
\begin{equation}
    \frac{d\rho}{dt} = -i\left[H,\rho\right] + \Gamma\left[S^{-}_c \rho S^{+}_c - \frac{1}{2}\left\{S^{+}_c S^{-}_c,\rho\right\}\right]
\end{equation}
where the Hamiltonian may be written as $H = J_zS^{z}_c$ and is identical to the one in Equation~\eqref{hamext}.

In this case, one can use the concept of coherent spin states, where spin states are associated with number states in quantum optics~\cite{radcliffe1971some}. Consider a collection of $S$ spins and define the fully polarized down state as $\ket{0}$,
\begin{equation}
    S^z_c \ket{0} = - S \ket{0}.
\end{equation}
The state with $p$ spin excitations can then be written as
\begin{equation}
    (S_c^+)^p \ket{0} = \sqrt{\frac{p!(2S)!}{(2S - p)!}} \ket{p},
\end{equation}
where $0 \leq p \leq 2S$, and $\ket{p}$ denotes the normalized Dicke state with $p$ excitations.

A spin coherent state can be defined as
\begin{equation}
    \ket{\chi} = \frac{1}{(1+|\chi|^2)^S}\exp(\chi S^{+}_c)\ket{0}.
\end{equation}
In the large-spin limit $S \rightarrow \infty$, the Holstein--Primakoff transform yields the approximate mapping~\cite{radcliffe1971some}
\begin{align}
    S^{-}_c &\rightarrow \sqrt{2S}\,a,\\
    S^{+}_c &\rightarrow \sqrt{2S}\,a^{\dagger},\\
    S^{z}_c &\rightarrow a^{\dagger}a - S,\\
    \chi &\rightarrow \frac{\alpha}{\sqrt{2S}},
\end{align}
under which the spin coherent state becomes the bosonic coherent state
\begin{equation}
    \ket{\alpha} = \exp\!\left(-\frac{1}{2}|\alpha|^2\right)\exp(\alpha a^\dagger)\ket{0}.
\end{equation}

In this limit, the collective-spin Hamiltonian $H = J_z S_c^z$ reduces to the harmonic-oscillator Hamiltonian $H \simeq J_z a^\dagger a$ (up to a constant), and the master equation for the collective decay channel becomes
\begin{equation}
    \frac{d\rho}{dt} = -iJ_z\left[a^{\dagger}a - \frac{L}{2}, \rho\right] + L\Gamma\left[a \rho a^{\dagger} - \frac{1}{2}\left\{a^{\dagger}a, \rho\right\}\right],
    \label{harmonicmaster}
\end{equation}
where $L = 2S$ is the total number of spins and $L\Gamma$ is the effective collective decay rate in the harmonic-oscillator description.

Assuming the system starts at a coherent state $\rho(0) = \ket{\alpha}\bra{\alpha}$, where $\ket{\alpha(t)} = e^{\alpha a^{\dagger} - \alpha^* a}\ket{0}$, we get (see the appendix~\ref{appendix}):
\begin{align}
    \dot{\alpha} = (-iJ_z - \frac{L}{2}\Gamma)\alpha \implies \alpha(t) = e^{(-iJ_z - \frac{L}{2}\Gamma)t} \alpha(0) \label{equationextreme1}
\end{align}
which shows that the decay rate indeed increases as the system size grows. Note that the coherent states in these approximations cover the all spins-up state when $|\alpha| \rightarrow \infty$, but do not approximate the singlet states that we have shown to decay slowly.

\section{Unravelings and Probability Dynamics for Davies Maps}
\label{section4}

Consider again Eq.~\eqref{LindbladEq}. The terms in $\mathcal{D}_{L_{k}}(\rho)$ play different roles in the evolution of the state. For example, if the system is a two level atom whose state is $\rho = \ket{\uparrow}\bra{\uparrow}$ and $L_k = \sigma_-$, then $\sigma_- \rho \sigma_+ = \ket{\downarrow}\bra{\downarrow}$, i.e. the system is projected onto the ground state. On the other hand $\sigma_+\sigma_- \ket{\uparrow}\bra{\uparrow} = \ket{\uparrow}\bra{\uparrow}$ and this operation only adds a phase to the state.

The Liouvillian may, thus, be written as:
\begin{equation}
    \frac{d\rho}{dt} = \mathcal{L}\rho = -i(H_{\text{eff}}\rho - \rho H_{\text{eff}}^{\dagger}) + \sum_k \lambda_k L_{k}\rho L_{k}^{\dagger},
    \label{LiouvillianSplit}
\end{equation}
where $H_{\text{eff}} = H - \frac{i}{2}\sum_k \lambda_k L_k^{\dagger} L_k$ is an effective non-Hermitian Hamiltonian. 

This separation into continuous evolution characterized by $H_{\text{eff}}$ and discontinuous "jumps" given by the second term allows the derivation of a stochastic method for the evolution of the master equation \cite{carmichael2009open,landi2024current}. The technique consists of evolving the state of the system for a given time step $\delta t$ with the non-Hermitian Hamiltonian and choosing whether a "jump" will occur \cite{molmer1993monte}. The total jump probability is defined as $dp = \delta t\sum_k \lambda_k \mathrm{Tr}[L_k \rho_c L_k^{\dagger}]$.

Explicitly, let $\rho_c(t)$ be the state of the system at time $t$ conditioned on a sequence of random jumps. To obtain the state at time $t + \delta t$, we choose a uniformly distributed random number $\epsilon \in [0, 1]$ and compare it with $dp$. The updated state $\rho_c(t + \delta t)$ is chosen as:
\begin{equation}
    \rho_c(t + \delta t) = 
    \begin{cases}
    \frac{L_k \rho_c(t) L_k^{\dagger}}{\mathrm{Tr}[L_k \rho_c(t) L_k^{\dagger}]} & \text{if } \epsilon < dp, \\
    \frac{\tilde{\rho}_c(t+\delta t)}{\mathrm{Tr}[\tilde{\rho}_c(t+\delta t)]} & \text{otherwise,}
    \end{cases}
    \label{propagation}
\end{equation}
where $\tilde{\rho}_c(t+\delta t)= \rho_c(t) - i \delta t (H_{\text{eff}}\rho_c(t) - \rho_c(t)H_{\text{eff}}^\dagger)$ is the unnormalized state under continuous evolution. If a jump occurs ($\epsilon < dp$), the specific jump operator $L_k$ is selected with probability $p_k = \lambda_k \mathrm{Tr}[L_k \rho_c L_k^{\dagger}] / \sum_j \lambda_j \mathrm{Tr}[L_j \rho_c L_j^{\dagger}]$.

The evolution of $\rho_c$ under a specific choice of Lindblad operators is called a \textit{quantum trajectory} of a particular \textit{unravelling} of the master equation. Importantly, the ensemble average over a large number of trajectories converges to the deterministic state evolution of the density matrix, i.e, $\rho(t) = \mathbb{E}(\rho_c(t))$.

In order to investigate the mechanisms behind the Mpemba effect, as outlined in Section~\ref{section2}, we first consider the model of a single qubit coupled to a thermal bath at zero temperature. The system Hamiltonian is given by
\begin{equation}
    H = \frac{1}{2}\omega\sigma^z,
\end{equation}
and the full Lindblad equation reads
\begin{equation}
    \frac{d\rho}{dt} = -i\left[H,\rho \right]  + \mu \mathcal{D}_{\sigma^{-}}(\rho).
\end{equation}

\begin{figure}[htbp]
    \subfloat[]{
        \includegraphics[width=0.75\linewidth]{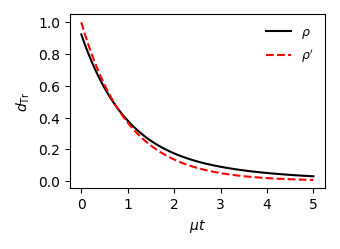}
        \label{subfig14}
    }
    \\
    \subfloat[]{
        \includegraphics[width=0.75\linewidth]{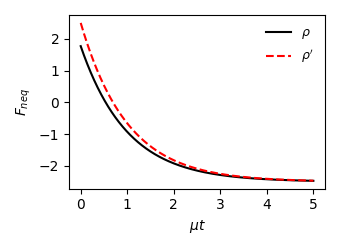}
        \label{subfig24}
    }
    \caption{Trace distance \protect\subref{subfig14} and non-equilibrium free energy \protect\subref{subfig24} for pure $\rho$ and $\rho^{\prime} = \ket{1}\bra{1}$. Here $\omega = 5$, $T=0\mu$, $\mu=1$. The reference state for $d_{\mathrm{Tr}}$ is $\rho_{ss} = \ket{0}\bra{0}$ and the initial state is determined by the randomly generated Bloch vector $\boldsymbol{R} = (0.0025964,-0.70710201,0.70710678)$.}
    \label{Unravelings1}
\end{figure}

Using the procedure given by Moroder \textit{et al.}~\cite{moroder2024thermodynamics} and outlined in Section \ref{section2}, we obtained the states that exhibit the QME. However, $F_{\mathrm{neq}}$ does not capture the effect of these states at zero temperature, since the crossing of the curves depends on how the von Neumann entropy varies. On the other hand, the trace distance ($d_{\mathrm{Tr}}(\xi,\rho_{ss})$ with $\xi = \rho \text{ , } \xi = \rho^{\prime} \text{ and } \rho_{ss} = \ket{0}\bra{0}$) still shows that $\rho^{\prime}$ reaches the equilibrium state faster (see Fig.~\ref{Unravelings1}). Thus, the trace distance allows us to study the system in its simplest form. To that end, we focus on the unravelings of the master equation for the two states.

\begin{figure}[htbp]
        \centering
        \includegraphics[width=0.75\linewidth]{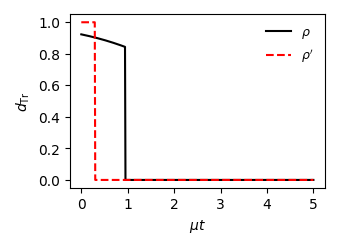}
    \caption{Single sample trajectory of the conditional trace distance for the states $\rho = \frac{1}{2}\left[\mathbb{I} + r_1\sigma^{x} + r_2\sigma^{y} + r_3\sigma^{z}\right]$ and $\rho^{\prime} = \ket{1}\!\bra{1}$. Here the reference state for $d_{\mathrm{Tr}}$ is $\rho_{ss} = \ket{0}\bra{0}$ and the parameters are $T=0\mu$, $\omega = 5$, $\mu = 1$ and time step $\delta t = 0.002\mu$. The figure shows the typical behavior for two pure states, where the Bloch vector is given by$\boldsymbol{R}= (0.0025964,-0.70710201,0.70710678)$ (generated randomly).}
    \label{subfig13}
\end{figure}

From the numerical results for single trajectories,  we observe that after the first jump the states are both in the steady state (Fig.~\ref{subfig13}), which, for this example, is the ground state. An intuition for the cause of the faster relaxation of $\rho^{\prime}$ may, then, be developed in terms of jump probabilities of both states. For pure states, $\rho^{\prime}$ is always given by the excited state (the highest-energy eigenvector of this Hamiltonian). Assuming $\rho$ is the density matrix for a general pure state $\ket{\psi}(t) = \alpha(t)\ket{0} + \beta(t)\ket{1}$, the probabilities $P_1$ and $P_1^{\prime}$ for a jump operator to act on the states in a time interval $\delta t$ are as follows:
\begin{equation}
    P_1^{\prime} = \mu \,\delta t,
    \qquad
    P_1 = \mu |\beta(t)|^{2}\delta t.
    \label{probabilities}
\end{equation}

Assuming $|\beta(t)|^2 < 1$, the probability that the first jump will occur for $\rho^{\prime}$ is always higher than for $\rho$. On the other hand, there is a probability $|\alpha(0)|^2$ that $\ket{\psi}$ will reach the ground state without a quantum jump. Assuming this is the case, the evolution of the state amplitudes follows~\cite{molmer1993monte}:
\begin{equation}
    \begin{aligned}
        \alpha(t) &= \alpha(0)\Big[|\alpha(0)|^2 + |\beta(0)|^2 e^{-\mu t}\Big]^{-\frac{1}{2}},\\
        \beta(t) &= \beta(0)\,e^{-\frac{\mu t}{2}}
        \Big[|\alpha(0)|^2 + |\beta(0)|^2 e^{-\mu t}\Big]^{-\frac{1}{2}}.
    \end{aligned}
\end{equation}
Thus, even if the quantum jump does not occur for $\ket{\psi}$, it will reach the ground state with an exponential rate $e^{-\mu t/2}$. This makes the survival probability, i.e., the probability that no jump occurs up to time $t$, behave as
\begin{equation}
    \begin{aligned}
        P_{s}^{\prime}(t) &= e^{-\mu t},\\
        P_{s}(t) &= |\alpha(0)|^2 + |\beta(0)|^2 e^{-\mu t}.
    \end{aligned}
\end{equation}
Thus,
\[
\frac{dP_s}{dt} = |\beta(0)|^2 \frac{dP_s^{\prime}}{dt},
\]
and the survival probability decays more slowly for $\ket{\psi}$. As a consequence, the survival probability for $\ket{\psi}$ is higher than that of $\ket{1}$ for all $\mu t>0$. Together with Eq.~\eqref{probabilities}, this guarantees that the averaged trajectories of the excited state reach the ground state faster than those of $\ket{\psi}$.

In the opposite regime $T \gg 0$, Another ingredient is required to understand the behavior of the two states. Since another jump operator now plays a role in the dynamics, the probabilities that the first jump occurs in a time interval $\delta t$ change to
\begin{equation}
    P_1^{\prime} = \mu(\bar{N} + 1)\delta t,
    \qquad
    P_1 = \mu(\bar{N} + |\beta(t)|^2)\delta t,
\end{equation}
and $\rho^{\prime}$ will, on average, undergo more jumps before the first jump of $\rho$.

Thus, the averaged $\rho^{\prime}$ over trajectories becomes more mixed faster and reaches thermalization more quickly.

\begin{figure}[htbp]
    \subfloat[]{
        \includegraphics[width=0.75\linewidth]{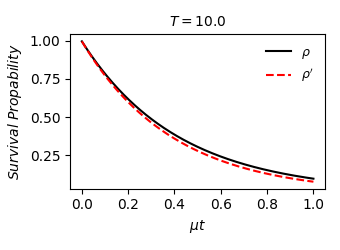}
        \label{subfigT10survival}
    }
    \\
    \subfloat[]{
        \includegraphics[width=0.75\linewidth]{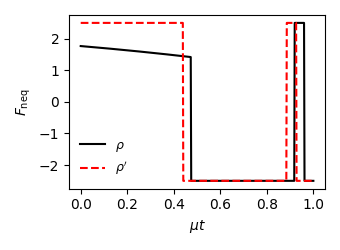}
        \label{subfigT10}
    }
\\
    \subfloat[]{
    \includegraphics[width=0.75\linewidth]{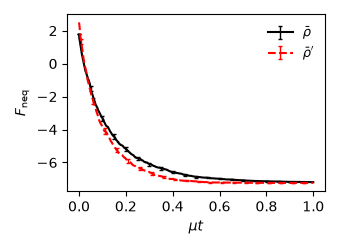}
    \label{averagedtrajec}
    }

    \caption{\protect\subref{subfigT10survival} Survival probability evolution for $\rho = \frac{1}{2}\left[\mathbb{I} + r_1\sigma^{x} + r_2\sigma^{y} + r_3\sigma^{z}\right]$ and $\rho^{\prime} = \ket{1}\bra{1}$, where both states are pure. \protect\subref{subfigT10} Single sample trajectory for the evolution of the non-equilibrium free energy for both states. \protect\subref{averagedtrajec}. $F_{neq}$ of the averaged state over trajectories. Here we used $\omega = 5$, $\mu = 1$, $T = 10\mu$, time step $\delta t= 0,002\mu$ and the state was averaged over 2000 trajectories. Consistency checks on the number of trajectories and time step show that the used parameters are adequate for convergence. The Bloch vector for $\rho$ is $\boldsymbol{R} = (0.0025964,-0.70710201,0.70710678)$ (generated randomly).}
    \label{singletrajpure}
\end{figure}

Returning to $F_{\mathrm{neq}}$, Fig.~\ref{singletrajpure} corroborates our description of the jump-probability dynamics underlying the quantum Mpemba effect. In these circumstances, the survival probabilities of $\rho^{\prime}$ and $\rho$ behave as (Fig.~\ref{subfigT10survival})
\begin{equation}
    \begin{aligned}
        P_{s}^{\prime}(t) &= \exp\!\left[-\mu(\bar{N} + 1)t\right],\\
        P_{s}(t) &= |\alpha(0)|^2 \exp\!\left[-\mu\bar{N}t\right]
        + |\beta(0)|^2\exp\!\left[-\mu(\bar{N} + 1)t\right].
    \end{aligned}
\end{equation}
Here, one finds that the survival probability changes as
\[\frac{dP_s}{dt}=
\left(\frac{\bar{N} (1 - \abs{\beta(0)}^2)}{\bar{N} + 1}e^{\mu t} + \abs{\beta(0)}^2\right)\frac{dP_s^{\prime}}{dt},
\]
and, as in the zero-temperature case, for $|\beta(0)|<1$ the survival probability for $\rho$ initially decays more slowly than for $\rho^{\prime}$. Since
\begin{equation}
    P_s - P^{\prime}_s = (1 - \abs{\beta(0)}^2)[e^{-\mu \bar{N} t} - e^{-\mu(\bar{N} + 1)t}] > 0.
\end{equation}
We see that the survival probability of $\rho$ will always be higher than that of $\rho'$ before thermalization. 

Note that the above results are not dependent on the specific unraveling chosen. Indeed, if we were to choose different Lindblad operators that characterize the same dissipative process, we would arrive at the same survival and jump probabilities. To see this define $L_1 = \sqrt{(\mu (\bar{N}) + 1)}\sigma^- $ and $L_2 = \sqrt{\mu \bar{N}}\sigma^+$. A new unraveling is obtained by the unitary transformation
\begin{align}
    \tilde{L}_1 &= \cos(\theta)L_1 + \sin(\theta)L_2\\
    \tilde{L}_2 &= -\sin(\theta)L_1 + \cos(\theta)L_2,
\end{align}
where $\theta$ is an arbitrary constant. By changing the master equation dissipators for $\mathcal{D}_{\tilde{L}_1}(\rho)$ and $\mathcal{D}_{\tilde{L}_2}(\rho)$ we arrive at the same results for jump and survival probabilities. 

For mixed states, it becomes interesting to look into the entropy averaged over trajectories, $\overline{S}_{\mathrm{vn}}(\rho(t))$. The von Neumann entropy is a highly non-linear quantity, and the average over trajectories generally leads to behavior that differs from the entropy of the averaged state, $S_{\mathrm{vn}}(\overline{\rho(t)})$. This allows the extraction of additional information from the system.

\begin{figure}[htbp]
    \subfloat[]{
        \includegraphics[width=0.75\linewidth]{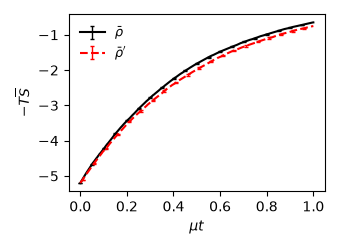}
        \label{stm1}
    }
    \\
    \subfloat[]{
        \includegraphics[width=0.75\linewidth]{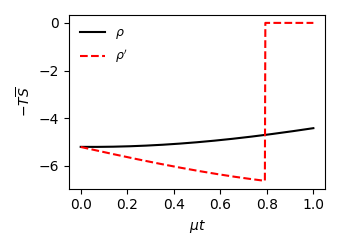}
        \label{stm2}
    }
    %
    %
    \\
    \subfloat[]{
        \includegraphics[width=0.8\linewidth]{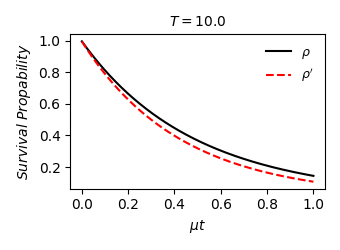}
        \label{stm3}
    }
    \caption{\protect\subref{stm1} Average over trajectories of $-TS$. Here we see a faster increase of the entropy for $\rho^{\prime}$, which is one of the factors leading the averaged state to remain mixed longer than $\rho$. \protect\subref{stm2} Single sample trajectory of the conditional non-equilibrium free energy for the states $\rho = \frac{1}{2}\left[\mathbb{I} + r_1\sigma^{x} + r_2\sigma^{y} + r_3\sigma^{z}\right]$ and $\rho^{\prime}$. \protect\subref{stm3} Survival probability evolution for both states. Here we used $\omega = 5$, $\mu = 1$, $T = 10\mu$,time step $\delta t= 0,002\mu$ and the entropy in \protect\subref{stm1} was averaged over 20000 trajectories. Consistency checks on the number of trajectories and time step show that the used parameters are adequate for convergence. The randomly generated Bloch vector was $\boldsymbol{R} = (0.52807291, 0.21585042, 0.02214326)$.}
    \label{singletrajecmixed}
\end{figure}

As seen in Fig.~\ref{stm1}, $T\overline{S}$ takes longer to reach its steady state for $\rho^{\prime}$. Looking at single trajectories (see Fig.~\ref{stm2}), one sees that one of the main factors contributing to the behavior of the averaged entropy is that the entropy of the state $\rho^{\prime}$ increases much faster than that of $\rho$ before the first jump. This is attributed to $\rho^{\prime}$ being diagonal in the energy eigenbasis~\cite{moroder2024thermodynamics}. The trajectories also show that, after the first jump, both states remain pure for the rest of the evolution.

The survival probabilities have a similar behavior as for pure states (Fig.~\ref{stm3}), since the initial state $\rho^{\prime}$ has a higher excited-state population by construction~\cite{moroder2024thermodynamics}. Here, coherences play an important role. We can see this by examining the explicit forms of the excited-state population, which we denote here as $\chi(t)$ to distinguish it from the pure-state amplitude and the squared coherences $|c(t)|^2$. Under the full thermal master equation, these evolve as follows:
\begin{equation}
    \begin{aligned}
        \chi(t) &= \chi^{\mathrm{eq}} + \big(\chi(0) - \chi^{\mathrm{eq}}\big)\exp\!\big[-\mu(2\bar{N}+1)t\big],\\
        |c(t)|^2 &= |c(0)|^2 \exp\!\big[-\mu(2\bar{N}+1)t\big],
    \end{aligned}
\end{equation}
where $\chi^{\mathrm{eq}}$ is the excited-state population of the thermal equilibrium state.

In particular, both the distance of the population to equilibrium and the coherences decay at the exact same temperature-dependent rate, $\Gamma = \mu(2\bar{N}+1)$. However, the presence of initial coherences in $\rho$ fundamentally alters the trajectory of the von Neumann entropy. The eigenvalues of the general density matrix, which dictate the entropy, are given by:
\begin{equation}
    \begin{aligned}
        \lambda_{\pm}(t) &=
        \frac{1 \pm \sqrt{1 - 4\det\rho(t)}}{2}\\
        &= \frac{1 \pm \sqrt{\big(2\chi(t) - 1\big)^2 + 4|c(t)|^2}}{2}.
    \end{aligned}
\end{equation}
For the state $\rho^{\prime}$, which is diagonal in the energy eigenbasis ($c^{\prime} = 0$), the entropy strictly follows the relaxation of the population. In contrast, for $\rho$, the non-zero coherences $|c(t)|^2$ add a strictly positive term under the square root, restricting entropy production for a longer portion of the evolution.

\begin{figure}[htbp]
    \subfloat[]{
        \includegraphics[width=0.75\linewidth]{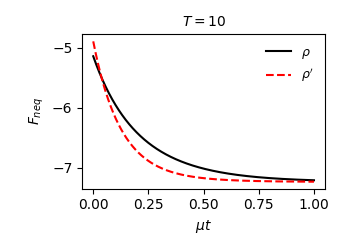}
        \label{stm12}
    }
    \\
    \subfloat[]{
        \includegraphics[width=0.75\linewidth]{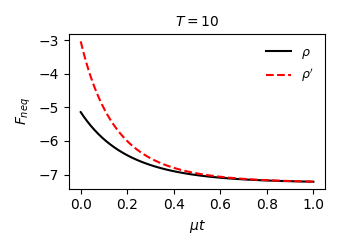}
        \label{stm22}
    }
    \caption{Comparison between different excited states $\rho^{\prime}$ with the same populations but different coherences (not strictly diagonal) and the original state $\rho$. \protect\subref{stm12} shows the results for the lower-coherence state, $\rho'_1$, in Eq.~\eqref{stateslab}, and \protect\subref{stm22} shows the results for the higher-coherence state, $\rho'_2$. Here we used $\omega= 5$, $\mu = 1$, $T = 10\mu$, and the randomly generated Bloch vector was $\boldsymbol{R} = (0.52807291, 0.21585042, 0.02214326)$.}
    \label{coherences}
\end{figure}

Because the von Neumann entropy
\[
S_{\mathrm{vn}} = -\big(\lambda_{+}\ln\lambda_{+} + \lambda_{-}\ln\lambda_{-}\big)
\]
is highly non-linear with respect to the density matrix eigenvalues, this restriction delays the entropy maximization for $\rho$. Therefore, $\rho^{\prime}$ experiences a faster initial entropy production rate and reaches the thermal state more quickly, despite starting with a higher non-equilibrium free energy. In this perspective, states with coherences $|c(0)|^2 \ll 1$ still thermalize faster as long as the excited state population is higher than that of the ground state. Figure~\ref{coherences} shows the comparison of the results for state evolution with the same $\rho$ as in Fig.~\ref{singletrajecmixed}. Here, we explicitly used the states
\begin{equation}
    \rho^{\prime}_1 =
    \begin{pmatrix}
        0.7 & 0.1\\
        0.1 & 0.3
    \end{pmatrix} \text{ and }
    \qquad
    \rho^{\prime}_2 =
    \begin{pmatrix}
        0.7 & 0.3\\
        0.3 & 0.3
    \end{pmatrix}
    \label{stateslab}
\end{equation}
instead of the usual $\rho' = U\rho U^{\dagger}$.

Thus, we can establish another criterion for determining whether two states will exhibit the QME. Here, given a density matrix $\rho$, we can slow down its approach to the equilibrium state by increasing its coherences. In this context, by choosing another state $\rho^{\prime}$ that has an excited state population sufficiently greater than that of $\rho$, alongside fewer coherences in the energy eigenbasis, the QME can be, in principle, engineered.

\section{Bosonic Gaussian Systems and a Microscopic Model}
\label{section5}

In order to further probe and understand the effect from a viewpoint of system--bath interaction, we consider a microscopic model in which we keep track of the bath state. In this context, we study a Gaussian bosonic system consisting of a single mode of an optical cavity (a single harmonic oscillator~\cite{adesso2014continuous}) coupled to a bath of harmonic oscillators. This is done using the covariance-matrix formalism, where the initial states of the system are chosen to be Gaussian and the states of each oscillator in the bath are thermal. The system is completely characterized by its covariance matrix $\boldsymbol{\sigma}$ and first moments $\boldsymbol{r}$~\cite{adesso2014continuous,landi2022nonequilibrium}, defined as

\begin{align}
    \boldsymbol{r} &= \langle \boldsymbol{R} \rangle,\\
    \sigma_{ij} &= \frac{1}{2}\langle R_i R_j + R_j R_i \rangle
    - \langle R_i \rangle \langle R_j \rangle,
\end{align}
where $\boldsymbol{R}$ is a $2N$-dimensional vector defined in terms of the quadrature operators,
\begin{align}
    \boldsymbol{R} &= (q_1,p_1,\dots,q_N,p_N),\\
    q_i &= \frac{1}{\sqrt{2}}(a_i + a^{\dagger}_i),
    \qquad
    p_i = \frac{1}{i\sqrt{2}}(a_i - a^{\dagger}_i).
\end{align}

The dynamics of a $N=2001$ mode system was then simulated. The first moments and covariance matrix obey the equations of motion~\cite{adesso2014continuous,landi2022nonequilibrium}
\begin{align}
    \frac{d\boldsymbol{r}}{dt} &= - W \boldsymbol{r},\\
    \frac{d\boldsymbol{\sigma}}{dt} &= -(W\boldsymbol{{\sigma}} + \boldsymbol{{\sigma}} W^{\dagger}),
\end{align}
where $W = -\Omega \boldsymbol{H}$, with $\boldsymbol{H}$ written in terms of $\boldsymbol{R}$, and $\Omega$ is the $N$-mode symplectic form
\begin{equation}
    \Omega = \bigoplus_{k = 1}^{N} \eta,
    \qquad
    \eta = \begin{pmatrix}
        0 & 1\\
        -1 & 0
    \end{pmatrix}.
\end{equation}
The first mode is taken as the system, and the remaining modes constitute the bath (initially in a thermal state).

The model adopted for the system-bath dynamics can be translated into the Hamiltonian
\begin{equation}
    \begin{aligned}
          H &= \omega \left( a^\dagger a + \frac{1}{2} \right) + \sum_{k=1}^{N_b} \epsilon_k \left( b_k^\dagger b_k + \frac{1}{2} \right) \\ 
          &+ \frac{J}{2} (a^\dagger + a) \sum_{k=1}^{N_b} (b_k^\dagger + b_k)
    \end{aligned}
  \end{equation}

where $a,b_k$ and their H.C. are bosonic annihilation and creation operators, respectively. J is the coupling strength between the system and bath modes, and $\epsilon_k$ is the frequency of the k-th bath mode. In order to obtain aproximate Markovian dynamics we set the system freequency to a number $\omega \gg J$. We also have used uniformly distributed frequencies for the bath modes around $\omega$.

\begin{figure}[htbp]
    \subfloat[]{
        \includegraphics[width=0.75\linewidth]{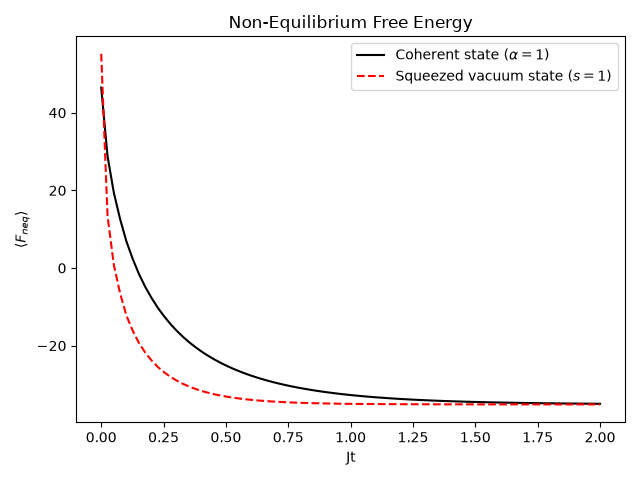}
        \label{sub1}
    }
    \vskip\baselineskip
    \subfloat[]{
        \includegraphics[width=0.75\linewidth]{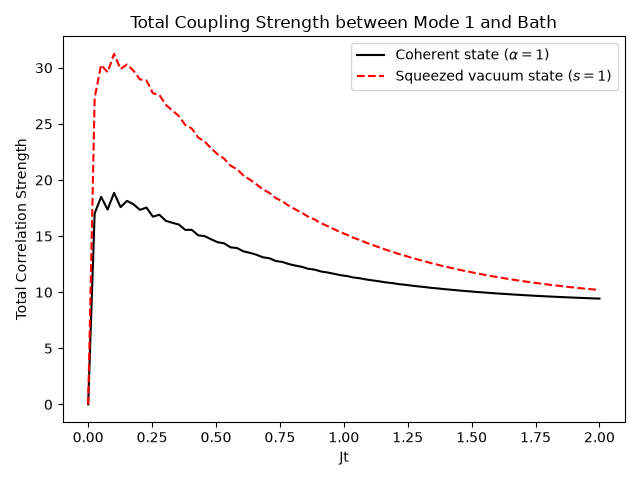}
        \label{sub2}
    }
    \vskip\baselineskip
    \subfloat[]{
    \includegraphics[width=0.9\linewidth]{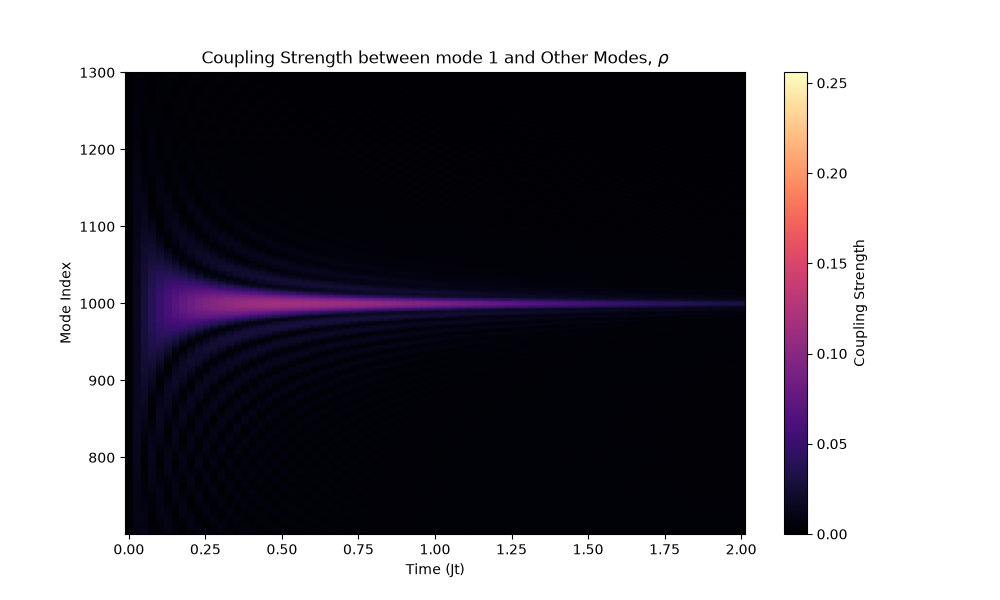}
        \label{sub3}
    }
    \vskip\baselineskip
    \subfloat[]{
        \includegraphics[width=0.9\linewidth]{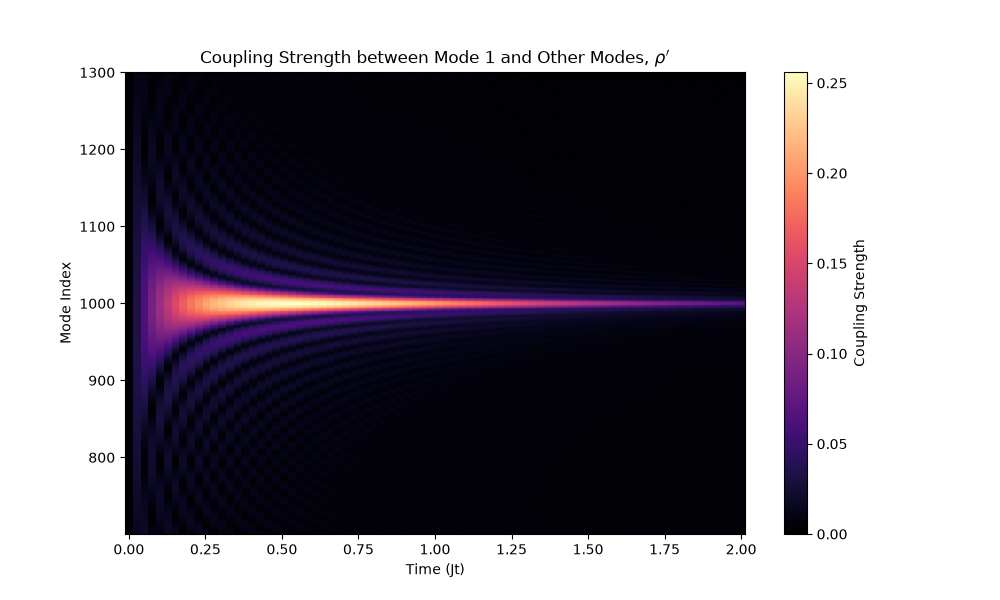}
        \label{sub4}
    }
    \caption{\protect\subref{sub1} Non-equilibrium free energy for the states $\rho = D(\alpha)\ket{0}\!\bra{0}D^{\dagger}(\alpha)$ and $\rho^{\prime} = S(s)\ket{0}\!\bra{0}S^{\dagger}(s)$ (coherent and squeezed vacuum states). \protect\subref{sub2} Total correlation strength between the system and the bath. \protect\subref{sub3}-\protect\subref{sub4} 2D plots showing the dynamics of the correlation strength between the system and each bath mode. Here we used  $T\approx 1.3\omega$, $J\approx 0.0016\omega$ and $\epsilon_k = 1 + 0.03k$, for $k \in \left[1,2000\right]$. Also the state parameters were $\alpha = 1$ and $s = 1$.}
    \label{ZenoMpemba3}
\end{figure}

The correlations between the system and bath modes are quantified through the Frobenius norm of the corresponding submatrices, which provides information about the correlation between the first and the $n$th mode (see Fig.~\ref{cov}). As was shown for the case of a single qubit in the previous section, coherences can play an important role in the thermalization of the system. Building on this idea, we can predict which states will present the QME based on their initial non-equilibrium free energy and the weights of their coherences in the energy eigenbasis. In order to obtain the quantum Mpemba effect, we select a coherent state and a squeezed vacuum state whose initial non-equilibrium free energy is higher than that of the former. The squeezed vacuum state tends to suppress odd excitations~\cite{gerry2023introductory} and, by choosing a squeezing parameter conviniently, we can obtain a state which has fewer coherences in the energy eigenbasis when compared with a coherent state. The results show that the effect is indeed observed for the chosen states (Fig.\ref{sub1}). 

In Fig.~\ref{sub2} we show the total correlation dynamics, defined as the sum of the Frobenius norms of all sub-blocks of $\boldsymbol{\sigma}$:
\begin{equation}
    \text{total correlation}
    = \sum_{i>1}\left(\|\sigma_{1,i}\|_F + \|\sigma_{i,1}\|_F\right).
\end{equation}
It is clear from the figure that the dynamics induce greater initial correlation build-up for the squeezed vacuum state,  which provides evidence that the effect is intrinsically connected to how strongly the bath interacts with a system for a given state. This build-up of correlations is similar across bath modes for the chosen states, as we show in Figs.~\ref{sub3} and~\ref{sub4}, with the exception of how strongly each state couples with the bath modes in time.

\begin{figure}
    \centering
    \tikzset{
        block matrix/.style = {
            matrix of math nodes,
            nodes in empty cells,
            left delimiter = [,
            right delimiter = ],
            every node/.append style = {
                outer sep = 0mm, inner sep = 0mm,
                text width = #1, text height = #1 * 0.60, text depth = #1 * 0.40,
                align = center, 
            }
        },
    }

    \begin{tikzpicture}
        \pgfdeclarelayer{back}
        \pgfsetlayers{back, main}

        \matrix [block matrix = 10mm] (M) 
        {
            & \sigma_{1, 2} & \dotsb & \sigma_{1, N} \\
            \sigma_{2, 1} & & & \\
            \vdots & & & \\
            \sigma_{N, 1} & & & \\
        };

        \begin{pgfonlayer}{back}
            \begin{scope}[opacity = 0.5]
                \fill [blue!40] (M-1-2.north west) rectangle (M-1-4.south east);
                \fill [blue!40] (M-2-1.north west) rectangle (M-4-1.south east);
                \fill [red!70 ] (M-1-1.north west) rectangle (M-1-1.south east);
            \end{scope}
        \end{pgfonlayer}

        \node[red!70 ] at ($(M-1-1.north)+(0, 0.5)$) {System};
        \node[blue!40] at ($(M-1-3.north)+(0, 0.5)$) {Correlations};
        \node[blue!40, rotate=90] at ($(M-3-1.west)+(-0.7, 0)$) {Correlations};

        \begin{scope}[dashed]
            \draw (M-1-2.north west) -- (M-4-2.south west);
            \draw (M-1-3.north west) -- (M-4-3.south west);
            \draw (M-1-4.north west) -- (M-4-4.south west);

            \draw (M-2-1.north west) -- (M-2-4.north east);
            \draw (M-3-1.north west) -- (M-3-4.north east);
            \draw (M-4-1.north west) -- (M-4-4.north east);
        \end{scope}
    \end{tikzpicture}

    
    \caption{Illustrative diagram of the covariance matrix and the blocks used to quantify the correlation strength between the system and the environment. In red is the 2x2 matrix which represents the mode considered to be the system. The blue blocks are the correlations between the first mode and the rest.}
    \label{cov}
\end{figure}

\section{Conclusion}

We have shown that one possible mechanism for obtaining the quantum Mpemba effect in Markovian open quantum systems is to exploit decoherence-free subspaces. This method has proven robust to changes in temperature and system size, and it opens new possibilities to utilize the effect in other areas, such as quantum optics~\cite{Scully2009,GrossHaroche1982} and ultracold gases~\cite{bloch2008many}. As a consequence of this mechanism, an extreme version of the quantum Mpemba effect can be achieved, characterized by a macroscopic speed-up toward the equilibrium state that scales linearly with system size. This has been shown both numerically and analytically (Section~\ref{section3} and Appendix~\ref{appendix}).

The quantum Mpemba effect has also been explored at the level of trajectories, revealing an interesting dynamics of jump probabilities which, when averaged, yields the effect. Here, the most interesting result is the dependence of the QME on the excited-state population. 

Furthermore, with a proposed microscopic model we kept track of the system--bath dynamics in a bosonic system, showing that the build up of larger initial correlations between system and bath for some states -- as opposed to others-- is intrinsically connected to the QME. 

It is instructive to compare the DFS-induced Mpemba effect with the traditional unitary transformation approach. The unitary method requires active, state-specific control to rotate a state into an optimal measurement basis. In contrast, the DFS mechanism shows that the Mpemba effect can emerge naturally from the structural symmetries of the environment (such as collective superradiant decay).

\begin{acknowledgments}
    This study was financed in part by the Coordenação de Aperfeiçoamento de Pessoal de Nível Superior – Brasil (CAPES) – Finance Code 001.
\end{acknowledgments}

\clearpage
\appendix

\section{Analytical Description of the Decay at Zero Temperature}
\label{appendix}

In order to solve this problem, we consider the system to start in a coherent state
$\rho(0) = \ket{\alpha}\bra{\alpha}$, assuming $\alpha$ to be time dependent:
\begin{equation}
    \ket{\alpha(t)} = D(\alpha(t))\ket{0}
    = e^{\alpha a^{\dagger} - \alpha^{*}a}\ket{0}.
\end{equation}
Using the identity $e^{A+B} = e^{A}e^{B}e^{\frac{1}{2}[A,B]}$, we can write
\begin{equation}
    D(\alpha(t)) = e^{-\frac{|\alpha|^2}{2}}e^{\alpha a^{\dagger}}e^{-\alpha^{*}a}.
\end{equation}
The following identities can then be obtained:
\begin{align}
    \frac{\partial D(\alpha)}{\partial \alpha}
    &= \left(a^{\dagger} - \frac{1}{2}\alpha^{*}\right)D(\alpha),\\
    \frac{\partial D(\alpha)}{\partial \alpha^{*}}
    &= -\left(a - \frac{1}{2}\alpha\right)D(\alpha),
    \label{identitiesD}
\end{align}
and thus, taking the time derivative of $\ket{\alpha}$ yields
\begin{equation}
    \begin{aligned}
         \frac{d}{dt}\ket{\alpha}
         &= \frac{\partial D(\alpha)}{\partial \alpha}\dot{\alpha}\ket{0}
         + \frac{\partial D(\alpha)}{\partial \alpha^{*}}\dot{\alpha}^{*}\ket{0}\\
         &= \Big[
         \dot{\alpha}\!\left(a^{\dagger} - \tfrac{1}{2}\alpha^{*}\right)
         - \dot{\alpha}^{*}\!\left(a - \tfrac{1}{2}\alpha\right)
         \Big]\ket{\alpha}.
    \end{aligned}
\end{equation}

The left-hand side of Eq.~\eqref{harmonicmaster} for a coherent state therefore reads
\begin{equation}
    \begin{aligned}
    \frac{d\rho}{dt}
    &= \Big[
        \dot{\alpha}\!\left(a^{\dagger} - \tfrac{1}{2}\alpha^{*}\right)
        - \dot{\alpha}^{*}\!\left(a - \tfrac{1}{2}\alpha\right)
      \Big]\rho \\
    &\quad + \rho
      \Big[
        \dot{\alpha}\!\left(a^{\dagger} - \tfrac{1}{2}\alpha^{*}\right)
        - \dot{\alpha}^{*}\!\left(a - \tfrac{1}{2}\alpha\right)
      \Big]^{\dagger},
    \end{aligned}
    \label{lefthandside}
\end{equation}

Using the defining property of a coherent state, namely that it is an eigenstate of
the annihilation operator, $a\ket{\alpha} = \alpha\ket{\alpha}$, the right-hand side of
Eq.~\eqref{harmonicmaster} becomes
\begin{equation}
    \begin{aligned}
        \frac{d\rho}{dt}
        ={}& -iJ_z\left(
            \alpha a^{\dagger}\ket{\alpha}\bra{\alpha}
            - \ket{\alpha}\bra{\alpha}\alpha^{*}a
        \right) \\
        &+ 2S\gamma\left(
            |\alpha|^{2}\ket{\alpha}\bra{\alpha}
            - \tfrac{1}{2}\alpha a^{\dagger}\ket{\alpha}\bra{\alpha}
            - \tfrac{1}{2}\ket{\alpha}\bra{\alpha}\alpha^{*}a
        \right).
    \end{aligned}
    \label{righthandside}
\end{equation}

Now, we note that
\begin{equation}
    \begin{aligned}
        D^{\dagger}(\alpha)a^{\dagger}D(\alpha)
        &= a^{\dagger}
        + [a^{\dagger},\alpha a^{\dagger} - \alpha^{*}a]
        + \dots \\
        &= a^{\dagger} + \alpha^{*},
    \end{aligned}
\end{equation}
where we used the Baker--Campbell--Hausdorff (BCH) formula%
\footnote{$e^{-A}Be^{A} = B + [B,A] + \frac{1}{2!}[[B,A],A] + \dots$}
and the fact that
\begin{equation}
    [[a^{\dagger},\alpha a^{\dagger} - \alpha^{*}a],
      \alpha a^{\dagger} - \alpha^{*}a]
    = [\alpha^{*},\alpha a^{\dagger} - \alpha^{*}a] = 0,
\end{equation}
since $\alpha^{*}$ is a complex number.

Using $D^{\dagger}(\alpha)D(\alpha) = \mathbb{I}$, we can express the action of
$a^{\dagger}$ on $\ket{\alpha}$ as
\begin{equation}
    \begin{aligned}
         a^{\dagger}\ket{\alpha}
         &= a^{\dagger}D(\alpha)\ket{0}
         = D(\alpha)\!\left(D^{\dagger}(\alpha)a^{\dagger}D(\alpha)\right)\ket{0} \\
         &= D(\alpha)(a^{\dagger} + \alpha^{*})\ket{0}
         = \alpha^{*}\ket{\alpha} + \ket{1_{\alpha}},
    \end{aligned}
    \label{adaggeralpha}
\end{equation}
where $\ket{1_{\alpha}} = D(\alpha)a^{\dagger}\ket{0}$.

Substituting Eq.~\eqref{adaggeralpha} into
Eqs.~\eqref{lefthandside} and~\eqref{righthandside}, we obtain
\begin{equation}
    \begin{aligned}
        \dot{\alpha}\ket{1_{\alpha}}\bra{\alpha}
        + \dot{\alpha}^{*}\ket{\alpha}\bra{1_{\alpha}}
        ={}& -iJ_z\left(
            \alpha^*\ket{\alpha}\bra{1_{\alpha}}
            - \alpha \ket{1_{\alpha}}\bra{\alpha}
        \right) \\
        &- S\gamma\left(
            \alpha\ket{1_{\alpha}}\bra{\alpha}
            + \alpha^*\ket{\alpha}\bra{1_{\alpha}}
        \right),
    \end{aligned}
\end{equation}
which leads directly to Eqs.~\eqref{equationextreme1}.

\clearpage

\bibliography{references}

\end{document}